\documentstyle[emulateapj]{article} 
\input{psfig.sty}
\singlespace
\def\kms{\,{\rm km\,s^{-1}}}

\def\msun{\,{\rm M_\odot}}

\def\etal{{et al.\ }}

\newcommand\beq{\begin{equation}}
\newcommand\eeq{\end{equation}}
\newcommand{\ba}{\begin{eqnarray}}
\newcommand{\ea}{\end{eqnarray}}
\def\gsim{\;\rlap{\lower 2.5pt
 \hbox{$\sim$}}\raise 1.5pt\hbox{$>$}\;}
\def\lsim{\;\rlap{\lower 2.5pt
   \hbox{$\sim$}}\raise 1.5pt\hbox{$<$}\;}
\def\spose#1{\hbox to 0pt{#1\hss}}
\newcommand{\lta}{\mathrel{\spose{\lower 3pt\hbox{$\mathchar"218$}}
      \raise 2.0pt\hbox{$\mathchar"13C$}}}
\newcommand{\gta}{\mathrel{\spose{\lower 3pt\hbox{$\mathchar"218$}}
      \raise 2.0pt\hbox{$\mathchar"13E$}}}

\lefthead{Volonteri, Madau, Quataert, \& Rees}
\righthead{MBH spin distribution}



\begin{document}
\submitted{ApJ, in press}

\title{The distribution and cosmic evolution of massive black hole spins}
\author{Marta Volonteri\altaffilmark{1}, Piero Madau\altaffilmark{1}, Eliot 
Quataert\altaffilmark{2}, \& Martin J. Rees\altaffilmark{3}}

\altaffiltext{1}{Department of Astronomy \& Astrophysics, University of 
California, Santa Cruz, CA 95064; pmadau@ucolick.org, marta@ucolick.org.}
\altaffiltext{2}{Department of Astronomy, 601 Campbell Hall, University of 
California at Berkeley, Berkeley, CA 94720; eliot@astron.berkeley.edu.}
\altaffiltext{3}{Institute of Astronomy, Madingley Road, Cambridge
CB3 0HA, UK; mjr@ast.cam.ac.uk.}

\begin{abstract}
We study the expected distribution of massive black hole (MBH) spins
and its evolution with cosmic time in the context of hierarchical
galaxy formation theories.  Our model uses Monte Carlo realizations of
the merger hierarchy in a $\Lambda$CDM cosmology, coupled to
semi-analytical recipes, to follow the merger history of dark matter
halos, the dynamics of the MBHs they host, and their growth via gas
accretion and binary coalescences. The coalescence of comparable mass
holes increases the spin of MBHs, while the capture of smaller
companions in randomly-oriented orbits acts to spin holes down.  We
find that, given the distribution of MBH binary mass ratios in
hierarchical models, binary coalescences alone do not lead to a
systematic spin-up or spin-down of MBHs with time: the spin
distribution retains memory of its initial conditions.  By contrast,
because of the alignment of a MBH with the angular momentum of the
outer accretion disk, gas accretion tends to spin holes up even if the
direction of the spin axis varies in time.  In our models, accretion
dominates over black hole captures and efficiently spins holes up. The
spin distribution is heavily skewed towards fast-rotating Kerr holes,
is already in place at early epochs, and does not change much below
redshift 5.  If accretion is via a thin disk, about 70\% of all MBHs
are maximally rotating and have radiative efficiencies approaching
30\% (assuming a ``standard'' spin-efficiency conversion). Even in the
conservative case where accretion is via a geometrically thick disk,
about 80\% of all MBHs have spin parameters $a/m_{\rm BH}>0.8$ and
accretion efficiencies $>$12\%. Rapidly spinning holes with high
radiative efficiencies may satisfy constraints based on comparing the
local MBH mass density with the mass density inferred from luminous
quasars (Soltan's argument).  Since most holes rotate rapidly at all
epochs, our results suggest that spin is not a necessary and
sufficient condition for producing a radio-loud quasar.
\end{abstract}


\section{Introduction}

Dynamical evidence indicates that massive black holes (MBHs) reside at
the centers of most nearby galaxies (Richstone \etal 1998). The tight
correlation observed between their masses and the stellar velocity
dispersion or mass of the host bulge (Gebhardt \etal 2000; Ferrarese
\& Merritt 2000; H\"{a}ring \& Rix 2004) suggests a close relationship
between the growth of black holes and spheroids in galaxy halos. It is
not yet understood which physical processes established this
correlation and how it is maintained through cosmic time with
such a small dispersion. 

Besides their masses, $m_{\rm BH}$,
astrophysical black holes are completely characterized by their spins,
$S=aGm_{\rm BH}/c$, $0\le a/m_{\rm BH}\le 1$. The spin of a MBH 
is expected to have a significant effect on its observational
manifestation.  For example, the spin determines the efficiency of
converting accreted mass into radiation and has implications for the
direction of jets in active nuclei (Rees 1984). The electromagnetic
braking of a rapidly spinning black hole may extract rotational energy
(Blandford \& Znajek 1977), convert it into directed Poynting flux and
electron-positron pairs, and power some radio galaxies and gamma-ray
bursts. The orientation of the spin is thought to determine the
innermost flow pattern of gas accreting onto Kerr holes (Bardeen \&
Petterson 1975). The coalescence of two spinning black holes in a radio
galaxy may cause a sudden reorientation of the jet direction, perhaps
leading to the so-called ``winged'' or ``X-type'' radio sources
(Merritt \& Ekers 2002).

The spin of a MBH is determined by the competition between a number of
physical processes.  Black holes forming from the gravitational
collapse of very massive stars endowed with rotation will in general
be born with non-zero spin (e.g. Fryer, Woosley, \& Heger 2002).  An
initially non-rotating hole that increases its mass by (say) 50\% by
swallowing material from an accretion disk may be spun up to $a/m_{\rm
BH}=0.84$ (Bardeen 1970). While the coalescence of two non-spinning
black holes of comparable mass will immediately drive the spin
parameter of the merged hole to $a/m_{\rm BH}\gta 0.8$ (e.g. Gammie,
Shapiro, \& McKinney 2004), the capture of smaller companions in
randomly-oriented orbits may spin down a Kerr hole instead (Hughes \&
Blandford 2003, hereafter HB).

In this paper we make a first attempt at estimating the distribution
of MBH spins and its evolution with cosmic time in the context of
hierarchical structure formation theories.  In the next section we
review the assembly of MBHs in popular cold dark matter (CDM)
cosmogonies.  We then address the evolution of MBH spin via
coalescences (\S3) and gas accretion (\S4).  We compute the spin
distribution of MBHs in \S5, and summarize our results and discuss
their implications in \S6.

\

\

\section{Assembly and growth of MBH\lowercase{s}}

The main features of a plausible scenario for the hierarchical
assembly, growth, and dynamics of MBHs in a $\Lambda$CDM cosmology
have been discussed in Volonteri, Haardt, \& Madau (2003), and
Volonteri, Madau, \& Haardt (2003).  Dark matter halos and their
associated galaxies undergo many mergers as mass is assembled from
high redshift to the present. The halo merger history is tracked
backwards in time with a Monte Carlo algorithm based on the extended
Press-Schechter formalism. ``Seed" holes form with intermediate masses
in the rare high-$\sigma$ peaks (``minihalos") collapsing at $z=20-25$
(Madau \& Rees 2001), and their growth and dynamical evolution is
followed in detail with a semi-analytical technique. The merging --
driven by dynamical friction against the dark matter -- of two
comparable-mass halo$+$MBH systems (``major mergers'') drags in the
satellite hole towards the center of the more massive progenitor,
leading to the formation of a bound MBH binary with separation of
$\sim$ pc. The long dynamical frictional timescales leave many MBHs
``wandering'' in galaxy halos after a merger between halos of large
mass ratio (``minor mergers'').

A two-component model for galaxy halos is adopted, where the dark
matter is distributed according to an NFW profile (Navarro, Frenk, \&
White 1997), while the initial central stellar distribution is a
singular isothermal sphere with stellar velocity dispersion,
$\sigma_*$. The latter is related to the halo circular velocity
following Ferrarese (2002). The present-day mass density of nuclear 
MBHs accumulates via gas accretion: in every major merger, the heavier hole
accretes at the Eddington rate a gas rest mass $\Delta m_0$. This
leads to a change in the total mass-energy of the hole given by
\begin{equation}
\Delta m=3.6\times 10^6\,\msun~V_{c,150}^{5.2},
\label{macc_eq}
\end{equation}
where $V_{c,150}$ is the circular velocity of the merged system in
units of 150 $\kms$ (Volonteri, Madau, \& Haardt 2003). The scaling
with the fifth power of the circular velocity of the host halo and the
normalization are fixed in order to reproduce the observed local
$m_{\rm BH}-\sigma_*$ relation. The quantities $\Delta m$ and $\Delta
m_0$ are related by $\Delta m=(1-\epsilon)\,\Delta m_0$, where
$\epsilon$ is the mass-to-energy conversion efficiency, equal for
thin-disk accretion to the binding energy per unit mass of a particle
in the last stable circular orbit.\footnote{The conversion from black
hole spin to radiative efficiency is still debated, and depends on
whether or not magnetic stresses exert a significant torque at the
last stable orbit (e.g. Krolik, Hawley, \& Hirose 2004); here we adopt
the standard definition for circular equatorial orbits around a Kerr
hole. For prograde orbits, the mass-to-energy conversion efficiency
$\epsilon\equiv 1-\hat E$ can then be expressed as $\hat
a=2/(3\sqrt{3})\left[2\sqrt{2/(2\epsilon-\epsilon^2)}-
(1-\epsilon)/(2\epsilon-\epsilon^2)\right]$.}\ This model was shown in
Volonteri, Haardt, \& Madau (2003) to reproduce well the observed
luminosity function of optically-selected quasars in the redshift
range $1<z<5$.

\subsection{MBH binaries}

The semimajor axis of a bound MBH binary continues to shrink owing to
dynamical friction from distant stars acting on each hole
individually, until the pair becomes ``hard'' when the separation
falls below $a_h=Gm_2(4\sigma_*^2)$ (Quinlan 1996), where $m_2(<m_1)$
is the mass of the lighter hole.  After this stage the MBH pair
hardens via three-body interactions, i.e., by capturing and ejecting
at much higher velocities the stars passing within a distance of order
the binary separation (Begelman, Blandford, \& Rees 1980).  Our scheme
assumes that the ``bottleneck'' stages of binary shrinking occur for
separations smaller than $a_h$; during a galactic merger, after a
dynamical friction timescale, we place the MBH pair at $a_h$ and let
it evolve. The hardening of the binary modifies the stellar density
profile, removing mass interior to the binary orbit, depleting the
galaxy core of stars, and slowing down further hardening.  The above
scheme can be modified by rare triple black hole interactions, when
another major merger takes place before the pre-existing binary has
had time to coalesce (Saslaw, Valtonen, \& Aarseth 1974). In this case
there is a net energy exchange between the binary and the third
incoming black hole, resulting in the ejection of the lighter hole and
the recoil of the binary. The binary also becomes more tightly bound.
Our scheme implicitly neglects the depopulation of the ``loss cone,''
i.e. it assumes a large supply of low-angular momentum stars (as one
might expect in significantly flattened or triaxial galaxies, Yu
2002). It also neglects the role of gaseous -- rather than stellar
dynamical -- processes in driving the evolution of a MBH binary
(e.g. Escala \etal 2004; Armitage \& Natarajan 2002).

If stellar dynamical and/or gas processes drive the binary
sufficiently close ($\lta 0.01$ pc), gravitational radiation will
eventually dominate angular momentum and energy losses and cause the
two MBHs to coalesce. 

\section{Spin change by binary coalescences}

HB have recently addressed how binary coalescence changes the spin of
the remnant hole. We follow here their treatment. The problem is
simpler for small mass ratios, $q\equiv m_2/m_1\ll1$, as the binary is
then well described by a test particle, the orbit of which shrinks due
to gravitational wave emission to the last stable orbit (LSO): the
smaller hole then plunges into the more massive one adding the LSO
angular momentum to its spin (since a black hole spin scales with its
mass squared, the small hole's spin can be neglected). While the test
particle description is not accurate for $q\gta 0.5$, binary
coalescences with mass ratios larger than 0.5 take place in only 10\%
of our simulations at $1<z<5$, and are even more rare at $z<1$ (see
Figure \ref{fig1}).  Moreover, the test particle description is
qualitatively correct in that $q \gta 0.5$ coalescences lead to
rapidly spinning holes most of the time (see Figs. \ref{fig2} and 
\ref{fig3}).

We adopt below geometrized units and set $c=G=1$; hatted quantities
have been made dimensionless by dividing out powers of mass,
e.g. $\hat a_1 \equiv a_1/m_1$. Before coalescence, the larger hole
has mass $m_1$ and spin $|{\bf S}|=\hat a_1m_1^2$. Kerr geodesics
orbits are specified by choosing their energy $E$, angular momentum
parallel to the spin, $L_z$, and ``Carter constant'' Q.  Orbits with
$Q\neq 0$ are inclined at angle $\theta$ with respect to the
equatorial plane, $\cos\theta=L_z/\sqrt{L_z^2+Q}\equiv \mu$.  

The smaller hole plunges into the large one carrying along its orbital
constants at the LSO\footnote{We have assumed here that the emission of
gravitational waves during the plunge phase does not affect the  
binary energy and angular momentum.  This may not be true for nearly equal 
mass mergers, for which there might not be a well-defined plunge
phase (e.g. Pfeiffer, Teukolsky, \& Cook 2000).}; given the inclination angle, 
one can determine the radius of the LSO, $L_z$, and $E$ by numerically solving the
system $R=\partial R/\partial r=\partial^2 R/ \partial r^2=0$, where
$R$ is the ``effective potential'' governing radial motion.  The
constants are bounded by their values for prograde ($\mu=1$) and
retrograde ($\mu=-1$) equatorial orbits,
\begin{eqnarray}
{\hat r} &\equiv & r_{\rm LSO}/m_1
\nonumber\\
&=& 3 + Z_2 \mp\sqrt{(3 - Z_1)(3 + Z_1 + 2 Z_2)},\\
\label{eq:rcirc_eq}
{\hat E} &\equiv & E_{\rm LSO}/m_2 =\left(1-{2\over 3 \hat r}\right)^{1/2},\\
\label{eq:Ecirc_eq}
{\hat L} &\equiv & L_{\rm LSO}/(m_2m_1) = \pm {2\over 3\sqrt{3}} [1+2(3\hat r 
-2)^{1/2}],\\
\label{eq:Lcirc_eq}
{\hat Q} & = & 0, 
\label{eq:Qcirc_eq}
\end{eqnarray}
where $Z_1$ and $Z_2$ are functions of $\hat a_1$ only (Bardeen, Press, 
\& Teukolsky 1972),
and the upper (lower) sign refers to prograde (retrograde) orbits. According
to HB, the numerical results are remarkably well fitted by 
\beq
\xi(\mu)\simeq |\xi_{\rm ret}|+{1\over 2}(\mu+1)(\xi_{\rm pro}-|\xi_{\rm ret}|), 
\eeq
where $\xi$ stands for $\hat r$, $\hat E$, or $\hat L$, and the ``ret'' and ``pro'' 
subscripts correspond to retrograde and prograde orbits, respectively. After 
capturing the smaller hole, the remnant MBH has mass and spin given by
\begin{eqnarray}
m' &=& m_1[1 + q {\hat E}({\hat a_1},\mu)]\;,
\label{eq:Mnew}\\
S'_z &=& m_1^2[{\hat a_1} + q {\hat L}_z({\hat a_1},\mu)]\;,
\label{eq:Sznew}\\
S'_\perp &=& q m_1^2{\hat L}_z({\hat a_1},\mu)\sqrt{\mu^{-2}-1}\;.
\label{eq:Sperpnew}
\end{eqnarray} 
The remnant hole is inclined at an angle $\Delta\theta$ relative to
the original hole, and has final spin parameter ${\hat a}'$:
\begin{eqnarray}
\Delta\theta &=& \arccos\left(S'_z/\sqrt{S^{\prime2}_z +
S^{\prime2}_\perp}\right)\;,
\label{eq:Dtheta}\\
{\hat a}' &=& {\sqrt{S^{\prime2}_z + S^{\prime2}_\perp}/m^{\prime2}}\;.
\label{eq:anew}
\end{eqnarray}

Figures \ref{fig2} and \ref{fig3} show how, following a single
coalescence, the spin of the remnant hole $\hat a'$ depends on the
mass ratio and inclination cosine for two representative values of the
initial spin $\hat a_1$. There is little change in spin for $q<0.05$,
while for larger mass ratios the angular momentum ${\bf L}$ at the LSO
overwhelms the initial spin ${\bf S}$ of the hole. For
randomly-distributed inclination angles, MBHs that are initially
rotating with $\hat a_1=0.6$ are spun up even further for $q>0.125$,
and are spun down otherwise. Rapidly rotating holes, with $\hat
a_1=0.9$, are typically spun down except for large mass ratios and
close-to-prograde orbits.

\section{Spin changes by accretion}

Accretion from a disk of gas orbiting the hole will cause the spin to
evolve in both magnitude and orientation. If the disk is in the {\it
equatorial} plane of a Kerr hole, and gas is dumped directly down the
hole from the LSO, then the accretion of a rest mass $dm_0$ will lead
to a change $dm=\hat E\,dm_0$ in the total gravitational mass and
$dS_z=\hat L_zm\,dm_0$ in the total angular momentum of the hole.  The
evolution of the hole spin is governed by the differential equation
\beq {d\hat a\over d\ln m} = {\hat L_z\over \hat E}-2\hat a.  \eeq
This was integrated by Bardeen (1970) to obtain the evolution law
\begin{eqnarray}
\hat a'&=& {\hat r^{1/2}\over 3} {m_1\over m'} \left[4-\left({3m_1^2\over 
m'^2}\hat r -2\right)^{1/2}
\right]~~{\rm for}~{m'\over m_1}\le \hat r^{1/2},
\nonumber\\
\hat a'&=& 1~~{\rm for}~{m'\over m_1}\ge \hat r^{1/2}.
\end{eqnarray} 
A hole that is initially non-rotating ($\hat r=6$) gets spun up to a maximally-rotating 
state ($\hat a'=1$) after a modest amount of accretion, $m'=m_1\sqrt{6}$.    
A maximally-rotating hole ($\hat a=1$) gets spun down by retrograde accretion ($\hat r=9$) 
to $\hat a'=0$ after $m'=m_1\sqrt{3/2}$. A $180^\circ$ flip of the spin of an extreme-Kerr 
hole ($\hat a=\hat a'=1$) will occur after $m'=3\,m_1$.    

What is the spin change induced by an accretion disk that is inclined
relative to the equatorial plane of the hole? Irrespective of the
infalling material's original angular momentum vector, Lense-Thirring
precession will impose axisymmetry close in, with the gas accreting on
either prograde or retrograde equatorial orbits.  Further out, the
disk will be warped where the transition from aligned to misaligned
flow occurs (Bardeen \& Petterson 1975). This suggests that accretion
of material with randomly oriented angular momentum vectors would lead
to {\it spin-down} of a MBH, given the larger LSO of retrograde orbits
(Moderski \& Sikora 1996).\footnote{This argument is similar to that
given by HB for why an ensemble of binary coalescences leads to
spin-down.}\ However, while the material falling into the hole is
aligned (or anti-aligned) with it, the torque that aligns the inner
disk with the hole must ultimately realign the hole with the outer
accretion disk (Rees 1978; Scheuer \& Feiler 1996), thus leading to
accretion via prograde equatorial orbits.  The timescale for this
alignment depends on how a warped accretion disk communicates the warp
and evolves in time, and is given roughly by (Scheuer \& Feiler 1996)
\beq t_{\rm align} \sim 3
{\hat a} \left({m_{\rm BH} \over \dot M}\right) \left(R_S \over
R_w\right)^{1/2} \left(\nu_1 \over \nu_2\right)
\label{align}
\eeq
where $\dot M$ is the accretion rate onto the hole, 
$R_S$ is the Schwarzschild radius, $R_w$ is the location of the warp,
and $\nu_1$ and $\nu_2$ are the viscosities associated with
dissipating motion in and out of the plane of the disk, respectively.
For a thin disk of vertical height $H$, sound speed $c_s$, and angular
velocity $\Omega$, $\nu_2/\nu_1 = (2\alpha^2)^{-1}$ for $1 \gg \alpha
> H/R$, where $\nu_1$ is given by the usual Shakura-Sunyaev
prescription $\nu_1 = \alpha c_s H = \alpha H^2 \Omega$ (Papaloizou \&
Pringle 1983).

The location of the warp in the disk ($R_w$) can be estimated by
equating the diffusion time of the disk's warp to the timescale for
Lense-Thirring precession, which implies (Natarajan \& Pringle 1998)
\beq 
{R_w \over R_S} \sim \left({\hat a} \over \sqrt{2}
\alpha\right)^{2/3} \left(R \over H\right)^{4/3} \left(\nu_1 \over
\nu_2\right)^{2/3}. \label{rwarp}
\eeq
Substituting equation (\ref{rwarp}) into equation (\ref{align}) yields
\beq t_{\rm align} \sim 3 \left({m_{\rm BH} \over \dot M}\right)
\left(\sqrt{2} {\hat a}^2 \alpha\right)^{1/3} \left(\nu_1 \over
\nu_2\right)^{2/3} \left(H \over R\right)^{2/3}. \label{align2} \eeq
Equation (\ref{align2}) shows that, for thin accretion disks with $H
\ll R$, the alignment time is much less than the timescale for the
mass of the hole to increase via accretion (using $\alpha$-disk
models, Natarajan \& Pringle 1998 and Natarajan \& Armitage 1999
estimate $t_{\rm align} \sim 10^5-10^6$ yrs for accretion near the
Eddington limit).  As a result, the hole will align itself with the
outer disk before it accretes much mass.  This implies that most of
the mass accreted by the hole will act to spin it up (i.e., the
magnitude of the spin increases in time), even if the direction of the
spin axis changes in time.  This is true unless the angular momentum
of the inflowing material changes significantly on a timescale $\ll
t_{\rm align}$.

If the mass supply rate to a MBH is super-Eddington, accretion
proceeds via a radiation pressure supported thick disk with $H \sim R$
(e.g., Begelman \& Meier 1982).  Equations (\ref{rwarp}) and
(\ref{align2}) show that in this case $R_w \sim R_S$ and alignment is
relatively inefficient, occurring on a timescale comparable to the
timescale for changes in the hole's mass (the Salpeter time).  It is
possible that much of the growth of MBHs at high redshift occurs via a
radiation pressure supported thick disk, rather than a thin accretion
disk.\footnote{If the mass supply rate is super-Eddington, most of the
inflowing gas is likely driven away in a radiation-pressure driven
outflow, limiting the accretion rate onto the hole to $\sim$ the
Eddington rate (e.g., Shakura \& Sunyaev 1973; Blandford \& Begelman
2004).  Thus accretion via a geometrically thick disk may still be
compatible with the mean accretion efficiency of $\sim 10\%$ inferred
by, e.g., Yu \& Tremaine (2002).}\ In fact, it would be surprising if
the mass supply rate were precisely $\sim$ Eddington (required for a
thin disk) during the entire growth of MBHs.  It is perhaps more
likely that the mass supply rate is initially larger in the dense
gas-rich environments of high redshift galaxies and then decreases
with time as the galaxy is assembled.

In the extreme case in which alignment is always relatively
inefficient, as would be expected for accretion via a thick disk, we
can generalize equations (\ref{eq:Mnew}), (\ref{eq:Sznew}), and
(\ref{eq:Sperpnew}) to determine the angular momentum evolution under
gas accretion as follows:
\begin{eqnarray}
dm &=& dm_0 \hat E(\hat a_1,\mu)\;,
\label{eq:Macc}\\
dS_z &=& m_1 dm_0 \hat L_z(\hat a_1,\mu)\;,
\label{eq:Szacc}\\
dS_\perp &=& m_1 dm_0 \hat L_z(\hat a_1,\mu)\sqrt{\mu^{-2}-1}\;.
\label{eq:Sperpacc}
\end{eqnarray} 
This equation can be integrated numerically to yield the final
orientation and magnitude of the hole spin after each accretion
episode, for a given inclination.


Finally, we note that Thorne (1974) showed that the radiation emitted
by the disk and swallowed by the hole produces a counteracting torque,
which prevents spin up beyond $\hat a'=0.998$.  Magnetic fields
connecting material in the disk and the plunging region may further
reduce the equilibrium spin by transporting angular momentum
outward. Fully relativistic magnetohydrodynamic simulations for a
series of thick accretion disk models show that spin equilibrium is
reached at $\hat a'\approx 0.93$ (Gammie, Shapiro, \& McKinney
2004). For simplicity, we will assume in the following that a MBH may
be spun up to a maximum equilibrium value of $\hat a'=0.998$.

\section{MBH spin distribution and evolution}

The combination of a halo merger tree and our semi-analytical scheme
to treat the growth of MBHs and their dynamics is a powerful tool for
tracking the evolution of MBH spins with cosmic time.  In our merger
tree, MBHs that undergo an accretion episode typically increase their
mass by about one e-folding. This is required in order to account for
the local mass density of MBHs with growth solely during major
mergers.  The distribution of fractional changes in hole mass from gas
accretion 
is shown in Figure \ref{fig4} for different redshift intervals. Note
how, at all epochs, a significant fraction of accretion events leads
to $\Delta m/m_1\gta 2-3$: these individual episodes will produce
rapidly-rotating holes independent of the initial spin.

To bracket the uncertainties and explore various scenarios we have run
different sets of Monte Carlo realizations. Our ``fiducial'' model
assumes seed holes are born with an initial spin of $\hat a=0.6$
(Fryer, Woosley, \& Heger 2002). The evolution (both in magnitude and
orientation) of MBH spins is driven by gas accretion and black hole
binary coalescences. Whenever a binary forms, the angle of
inclination of the smaller hole relative to the equatorial plane is
chosen randomly from an isotropic distribution.  We assume that
during a major merger the gas accretes via a thin disk; efficient
alignment between the hole and the outer disk implies that most of the
mass is accreted in prograde equatorial orbits (\S 4).  

Figure \ref{fig5} shows the ensuing spin distribution of MBHs at
different epochs. The left panel depicts the effect of binary
coalescences alone on the spin distribution.  At late epochs binaries
with small mass ratios are common (Fig. \ref{fig1}), as some less
massive holes that were wandering in galaxy halos at high redshift --
with an orbital decay timescale comparable to the Hubble time --
finally find their way to galaxy centers. In addition, our assumption
that, following a major merger between halos, a new accretion episode
is triggered only on the more massive hole, probably somewhat
underpredicts the mass ratio of coalescing holes. A more realistic
model in which both holes accrete would lead to somewhat larger mass
ratios, though this effect is likely to be mild given that the mass
accreted per merger is comparable to the mass of the pre-existing hole
(Fig. \ref{fig4}).

Figure \ref{fig5} shows that repeated captures of smaller holes cause
a spread both towards $\hat a>0.6$ (spin up) and $\hat a<0.6$ (spin
down) in $N(\hat a)$. The relatively flat distribution of $N(q)$ for
captures in hierarchical models (Fig. \ref{fig1}) implies that neither
spin-up nor spin-down is particularly favored.  The holes random walk
around the initial ``seed'' value, and the spin distribution retains
memory of the initial rotation black holes are endowed with at
birth. For comparison, we also show a case in which seed holes are
born {\it non-spinning} instead (dashed histogram in the left
panel). Again, at all redshifts, the spin distribution remains peaked
around $\hat a=0$.  The right panel in Figure \ref{fig5} includes spin
changes both by captures and by gas accretion. Gas accretion dominates
the spin evolution over coalescences (the initial mass in seed holes
is a tiny fraction of the mass density observed today).  At all
epochs, nearly all MBHs are spun-up by accretion to $\hat a \approx
1$.

Figure \ref{fig6} shows the spin evolution history of two MBHs that
end up in massive halos with $M_h=10^{12}\,\msun$ at $z=0$.  Note how
gas accretion (lower panel) efficiently spins the holes up, while
binary coalescences lead to both spin-up and spin-down.
Figures \ref{fig7} and \ref{fig8} explore the uncertainties introduced
by some of our assumptions. Figure \ref{fig7} shows the evolution of
the spin under binary coalescences alone, comparing two examples where
we placed seed holes in $3 \sigma$ and $3.5 \sigma$ peaks at high
redshift.  The former example has a total of $\sim 50$ times more seed
holes, leading to $\sim 10$ times more binary
coalescences;\footnote{The nonlinear increase in the number of binary
coalescences in going from 3.5 to 3 $\sigma$ peaks is due to several
factors: (1) there are a larger number of halo minor mergers that
never form MBH binaries because of the long dynamical friction
timescales, and (2) a larger number of holes are ejected from their
host galaxies by triple interactions and gravitational recoil.}\
the total mass in seed holes is now a larger fraction of the mass
density in MBHs observed today.  Despite the significant increase in
the number of binary coalescences, the spin evolution is reasonably
similar in the two examples.  The most notable difference is that
there are fewer rapidly rotating holes at low redshift in the case
with more seed holes, because of spin-down by capture of smaller
companions at late times (see Fig. \ref{fig1}).  The overall effect
is, however, rather mild, and the spin distribution still retains
significant memory of the initial conditions; binary coalescences
alone thus do not lead to a significant systematic spin-up or
spin-down of MBHs.

Figure \ref{fig8} shows an example where alignment of the hole and the
outer disk is inefficient, as would be expected if accretion is via a
geometrically thick disk (\S4).  We assume that the initial
orientation between the black hole's spin and the accretion disk
rotation axis is random, and integrate equations
(\ref{eq:Macc})-(\ref{eq:Sperpacc}) for every accretion episode,
i.e. we treat binary coalescences and gas accretion in a similar way.
This would seem to favor spin-down of the MBHs, and yet even in this
case most of the holes are rotating quite rapidly at all epochs.  The
reason is that the accreted mass is typically larger than the hole's
mass (Fig. \ref{fig4}).  Most individual accretion episodes thus
produce rapidly-rotating holes independent of the initial spin.

\section{Summary and discussion}

We have computed the expected distribution of MBH spins and its
evolution with cosmic time in the context of hierarchical structure
formation theories. A subset of the current authors have previously
described models for the birth, growth, and dynamics of MBHs that
traces their build-up far up the dark halo ``merger tree" in a
$\Lambda$CDM cosmology (see, e.g., Volonteri et al. 2003).  Here we
have extended this work to follow the combined effects of black
hole-black hole coalescences and accretion from a gaseous disk on the
magnitude and orientation of MBH spins.

\begin{figure}
\plotone{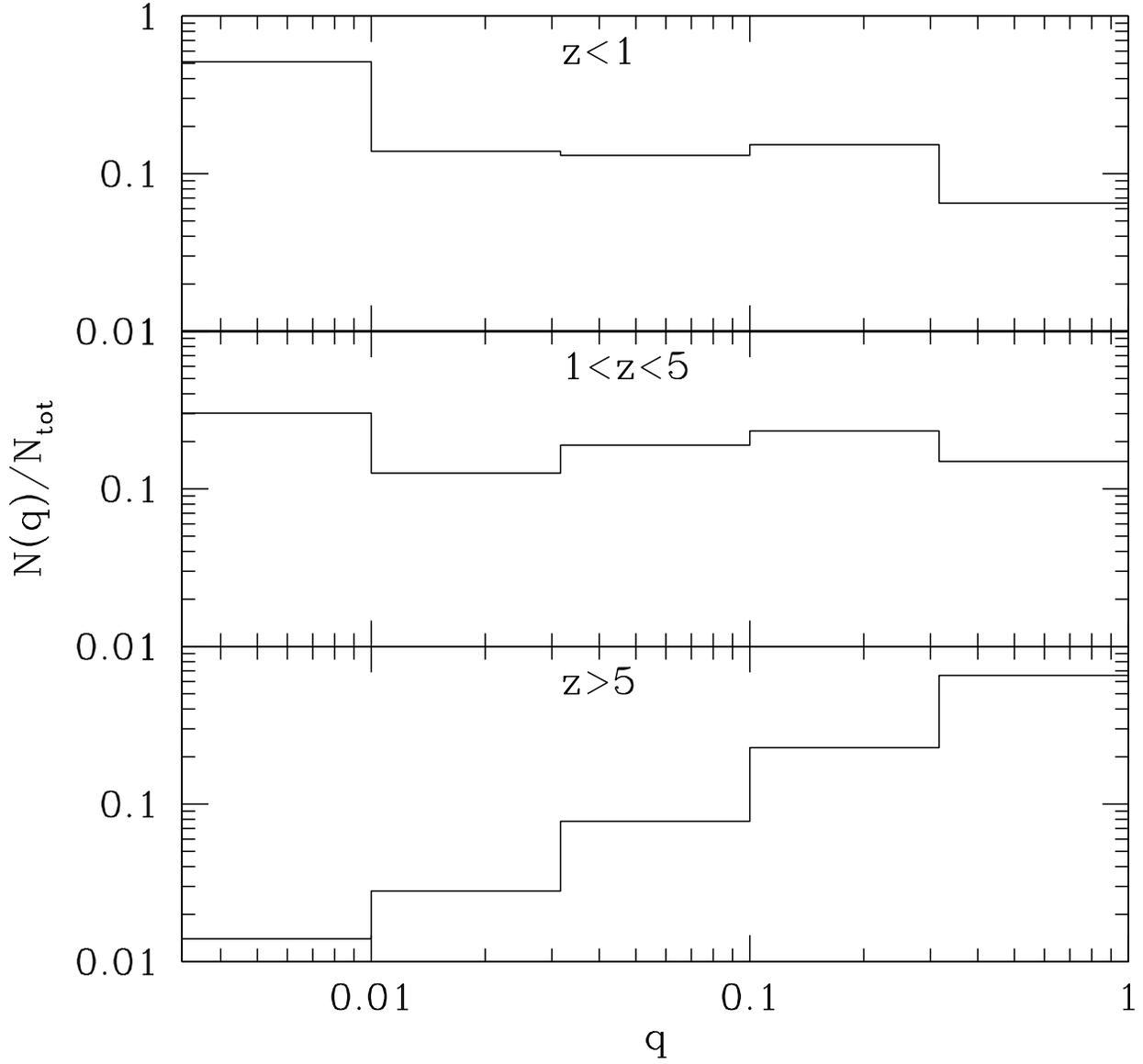} 
\caption{\footnotesize Normalized distribution of mass ratios of
coalescing MBH binaries at three different epochs. Note that at low
redshift MBHs typically capture much smaller companions.}
\label{fig1}
\end{figure}

\begin{figure}
\plotone{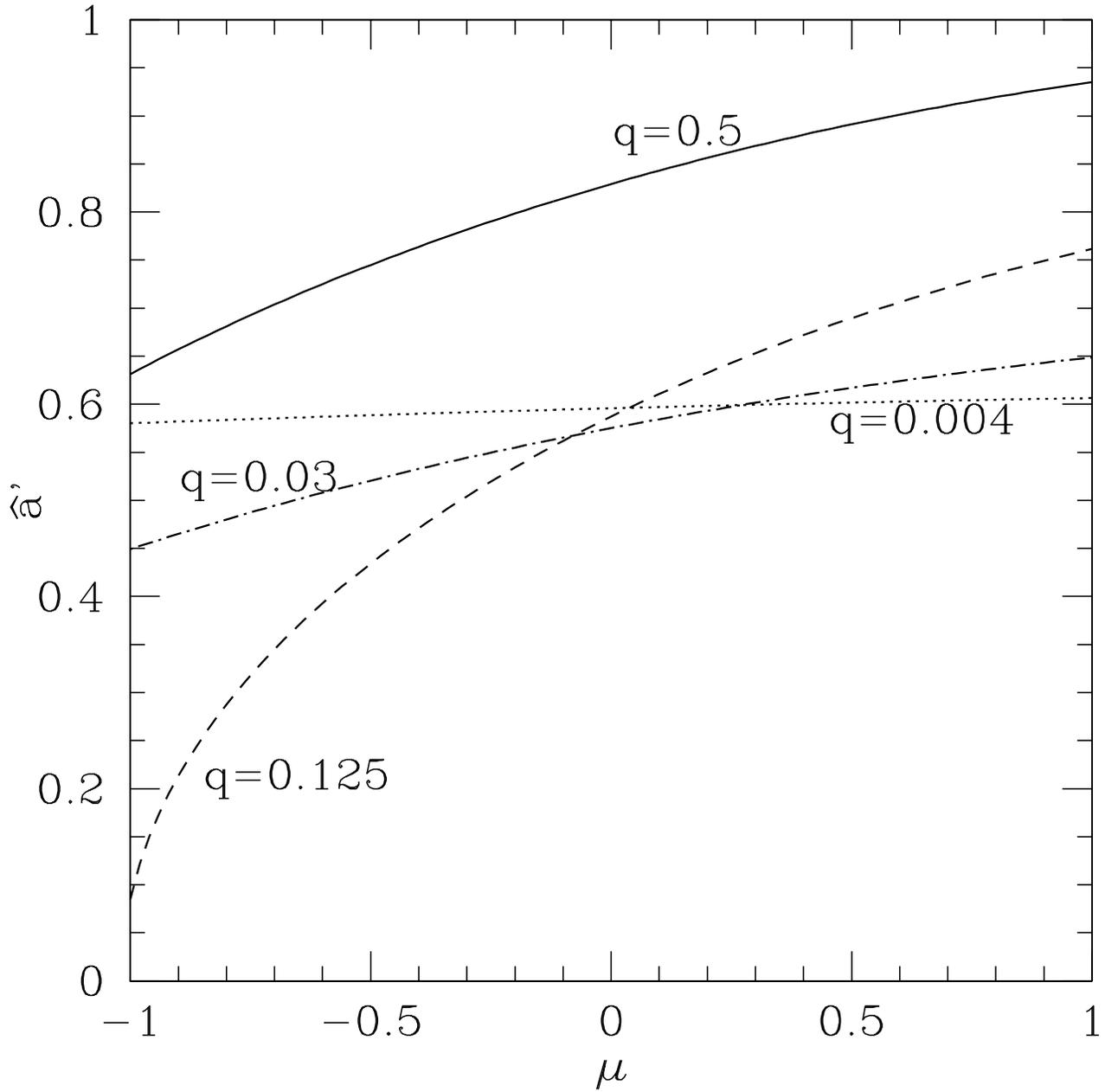}
\caption{\footnotesize Final spin $\hat a'$ of the remnant MBH as a function of 
inclination angle $\mu=\cos \theta$, for different binary mass ratios $q=m_2/m_1$. 
Before coalescence, the larger hole has spin $\hat a_1 =0.6$. 
The remnant is typically spun down for $q<0.125$, and is spun up otherwise.}
\label{fig2}
\end{figure}

\begin{figure}
\plotone{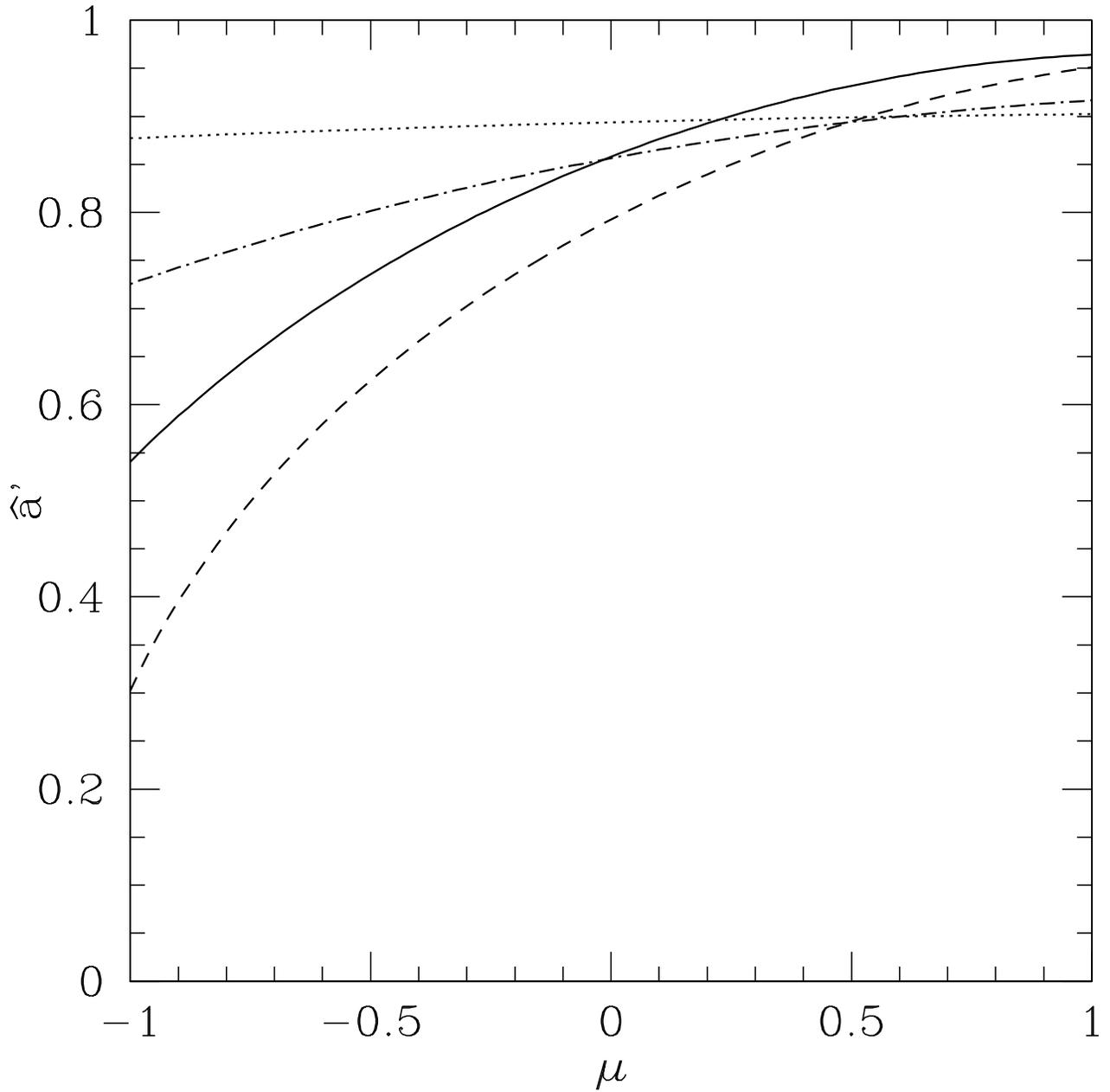}
\caption{ Same as Fig. \ref{fig2}, but with $\hat
a_1=0.9$. For most of the parameter space, the remnant MBH is spun down.}
\label{fig3}
\end{figure}

\begin{figure}
\plotone{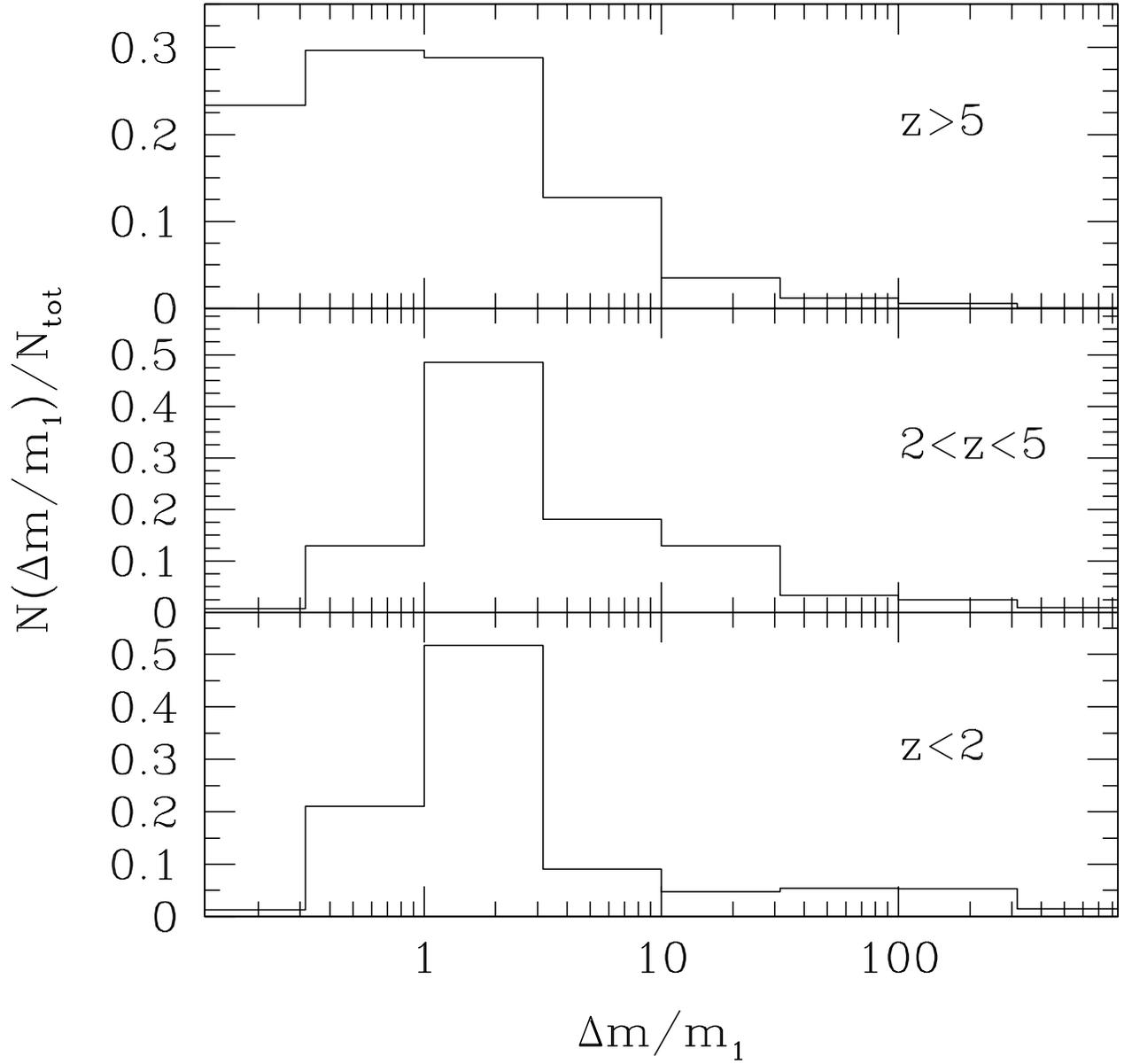}
\caption{ Distribution of fractional changes in the mass of a MBH 
as a result of the accretion of material after a major halo merger. Different 
redshift intervals are shown.
}
\label{fig4}
\end{figure}

\begin{figure}
\plotone{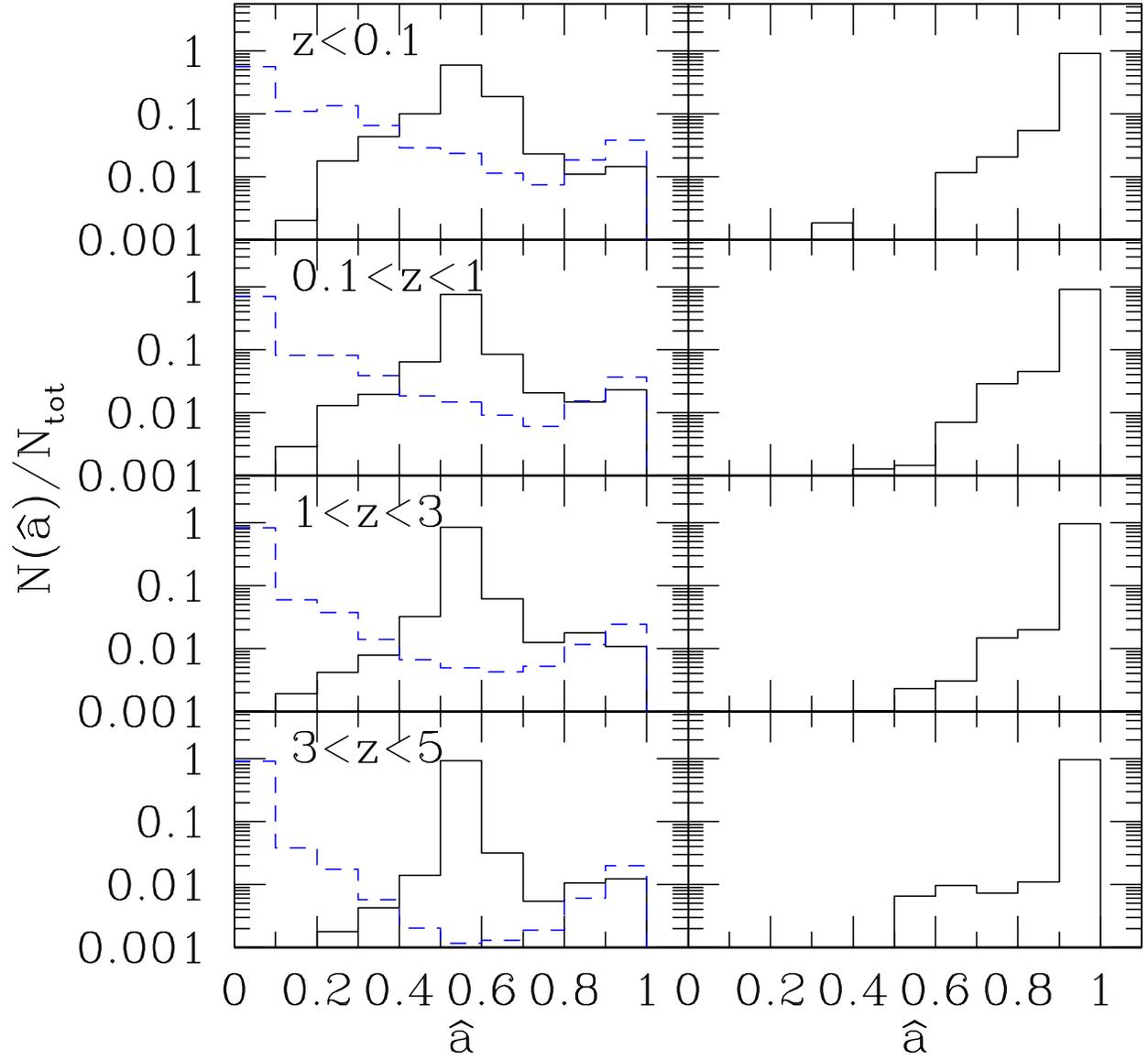}
\caption{ Distribution of MBH spins in different redshift intervals.
{\it Left panel:} effect of black hole binary coalescences only. 
{\it Solid histogram:} ``seed'' holes are born with $\hat a=0.6$.
{\it Dashed histogram:} ``seed'' holes are born non-spinning.
{\it Right panel:} spin distribution from binary coalescences and gas accretion.
``Seed'' holes are born with $\hat a=0.6$, and are efficiently spun up
by accretion via a thin disk.
}
\label{fig5}
\end{figure}

\begin{figure}
\plotone{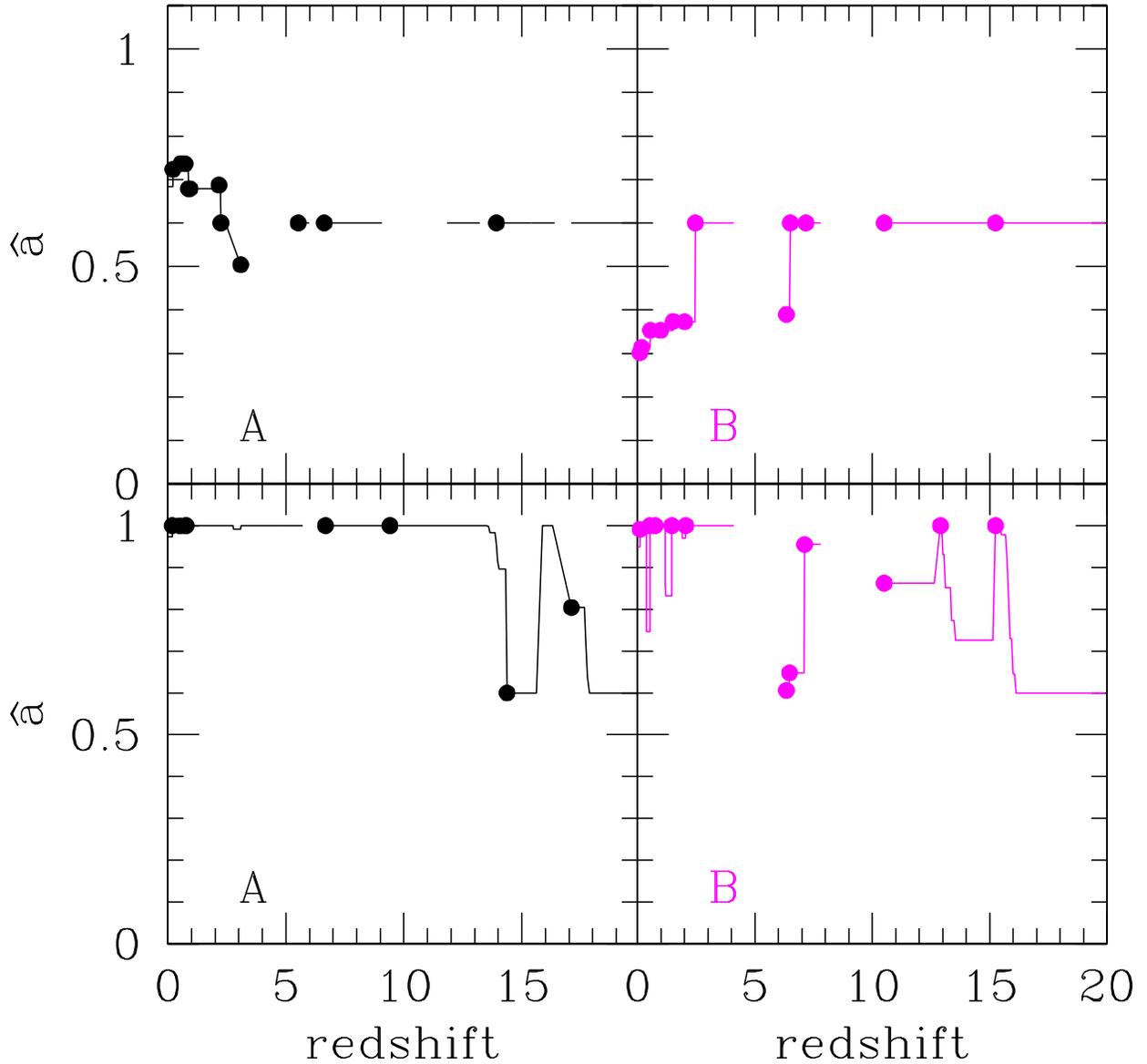}
\caption{ The spin evolution of two MBHs, ``A" ({\it left
panels}) and ``B''({\it right panels}), that end up in massive halos
with $M_h=10^{12}\,\msun$ at $z=0$. The initial spin of ``seed'' holes
is $\hat a=0.6$.  {\it Upper panel}: the spin of black holes is
modified by binary coalescences only. {\it Lower panel}: the spin
evolution is driven by black hole binary coalescences and gas
accretion.  The dots mark the epoch when two MBHs coalesce.  Note the
intermittent times when halos are devoid of a central MBH, after an
ejection due to radiation recoil. The halo (re)gains a central MBH
after the original one falls back in or the halo experiences a merger
with another halo, whose MBH becomes the central hole of the newly
formed galactic system.}
\label{fig6}
\end{figure}

\begin{figure}
\plotone{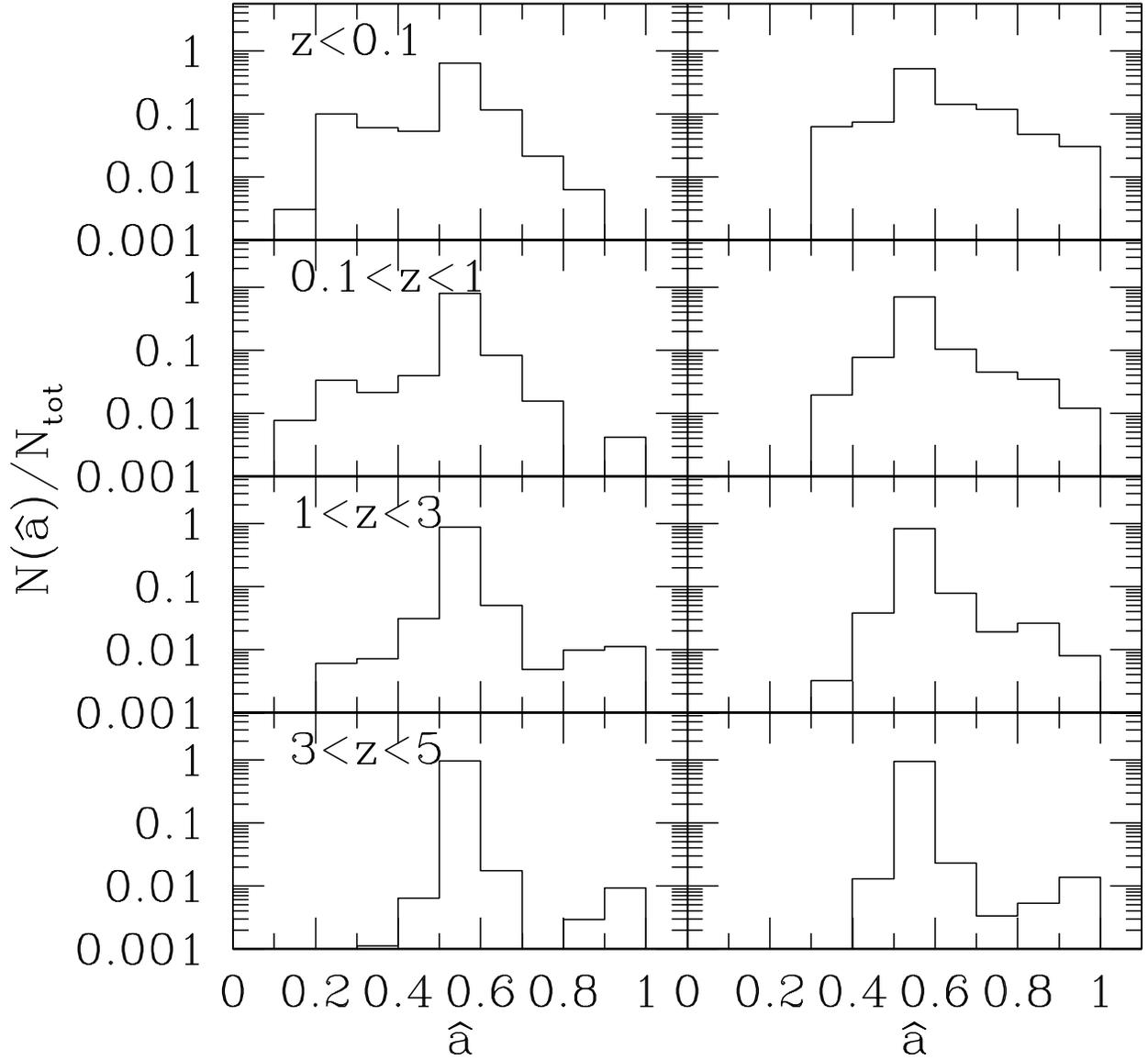}
\caption{\footnotesize The spin distribution for MBHs that end up in massive 
halos with $M_h=10^{12}\msun$ at $z=0$. Spins evolve under by binary coalescences
alone.  {\it Left panel}: ``seed'' holes are placed in $3 \sigma$
peaks at $z=20$.  {\it Right panel:} ``seed'' holes are placed in $3.5
\sigma$ peaks at $z=20$.  In both cases black holes are born with
$\hat a=0.6$.  The model in the left panel has $\approx 10$ times more
coalescences than that in the right panel, but the resulting spin
distributions are quite similar.}
\label{fig7}
\end{figure}

We find that binary coalescences cause no significant systematic
spin-up or spin-down of MBHs: because of the relatively flat
distribution of MBH binary mass ratios in hierarchical models
(Fig. \ref{fig1}), the holes random-walk around the spin parameter
they are endowed with at birth, and the $N(\hat a)$ distribution
retains significant memory of the initial rotation of ``seed'' holes
(Figs. \ref{fig5} and \ref{fig7}).

\begin{figure}
\plotone{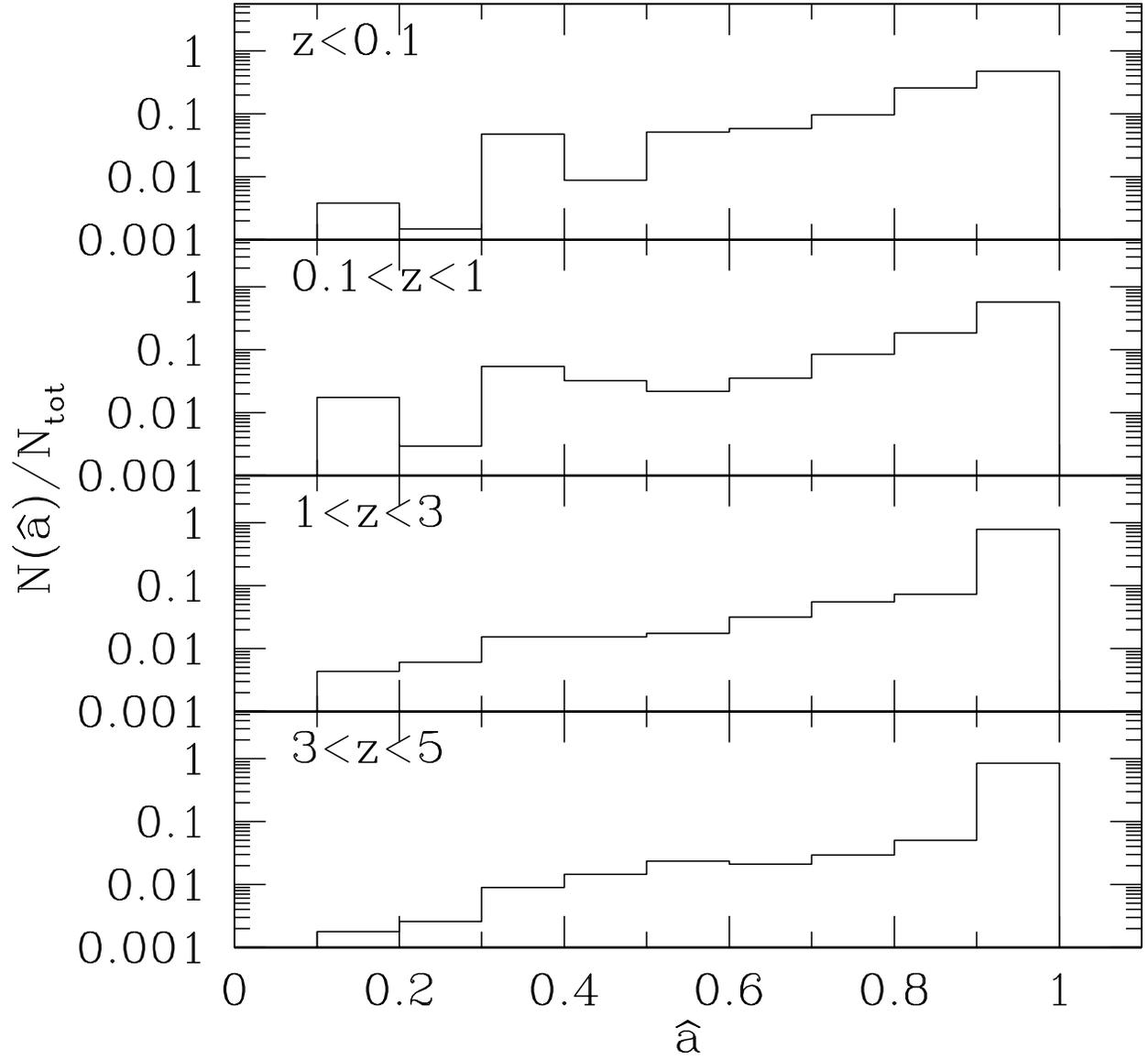}
\caption{\footnotesize Distribution of MBH spins in different redshift
intervals.  The spin is modified by binary coalescences and by
accretion; alignment of the hole and the outer disk is assumed to be
inefficient, as would be expected for accretion via a geometrically
thick disk. The initial orientation between hole's spin and disk
angular momentum is assumed to be random.}
\label{fig8}
\end{figure}

In our models, accretion, not binary coalescences, dominates the spin
evolution of MBHs.  Accretion can lead to efficient spin-up of MBHs
even if the angular momentum of the inflowing material varies in
time. This is because, for a thin accretion disk, the hole is aligned
with the outer disk on a timescale that is much shorter than the
Salpeter time (eq. \ref{align2}; Natarajan \& Pringle 1998), leading
to accretion via prograde equatorial orbits.  As a result, most of the
mass accreted by the hole acts to spin it up, even if the orientation
of the spin axis changes in time.  For a geometrically thick disk,
alignment of the hole with the outer disk is much less efficient,
occurring on a timescale comparable to the Salpeter time.  In this
case we still find that most holes are rotating rapidly
(Fig. \ref{fig8}).  This is because, in any model in which MBH growth
is triggered by major mergers, every accretion episode must typically
increase a hole's mass by about one e-folding to account for the local
MBH mass density and the $m_{\rm BH}-\sigma_*$ relation.  Most
individual accretion episodes thus produce rapidly-rotating holes
independent of the initial spin.

\begin{figure}
\plotone{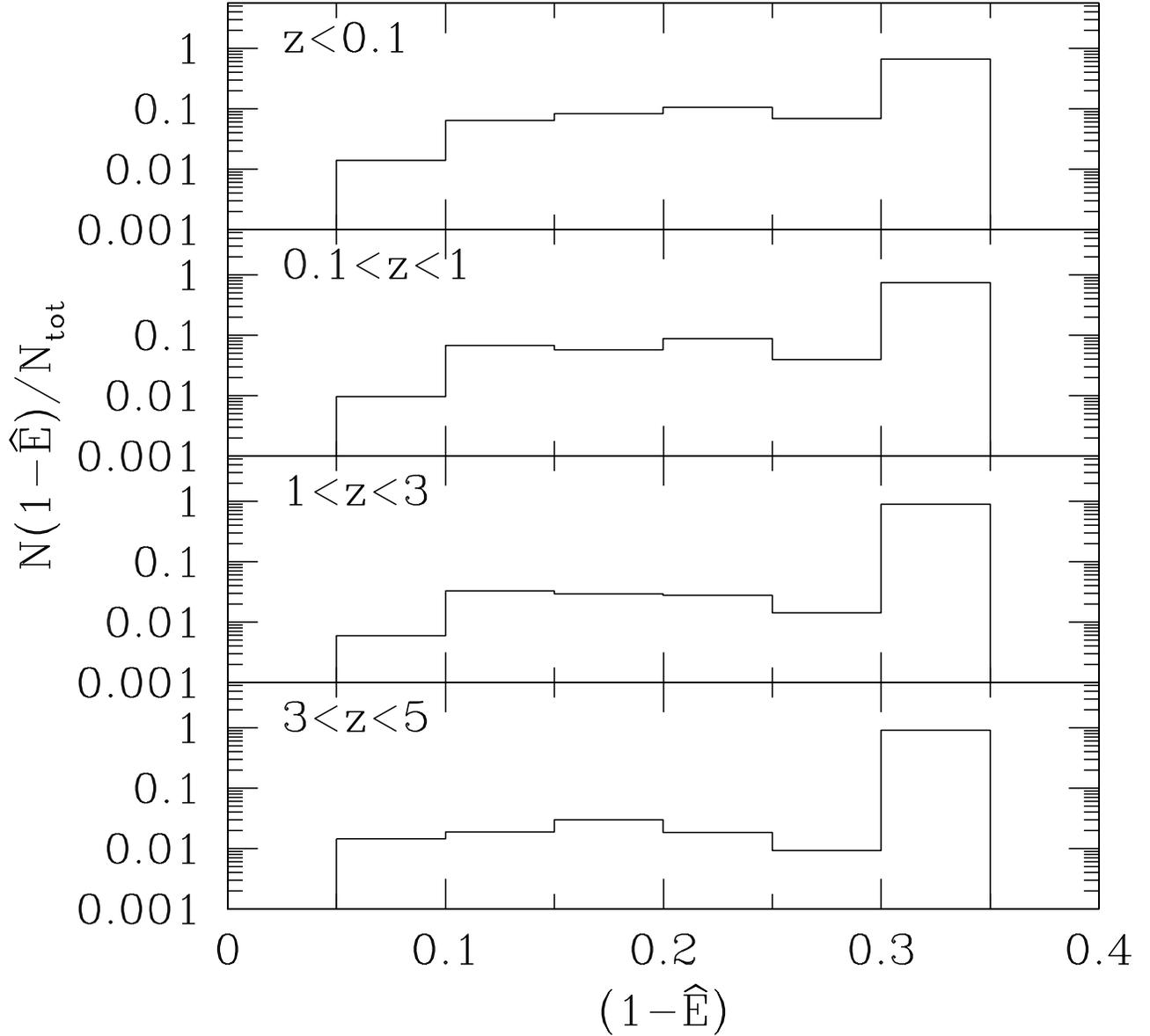}
\caption{ Distribution of accretion efficiencies,
$\epsilon\equiv 1-\hat E$, in
different redshift intervals (assuming that the energy radiated is the
binding energy at the LSO). The spin distribution from binary
coalescences and gas accretion has been calculated assuming the holes
accrete via a thin disk on prograde equatorial orbits.}
\label{fig9}
\end{figure}

Under the combined effects of accretion and binary coalescences, we
find that the spin distribution is heavily skewed towards
fast-rotating Kerr holes, is already in place at early epochs, and
does not change significantly below redshift 5.  As shown in Figure
\ref{fig9}, about 70\% of all MBHs are maximally rotating and have
mass-to-energy conversion efficiencies approaching 30\%. 
Note that if the equilibrium spin attained by accreting MBHs is lower than
the value of $\hat a = 0.998$ used here, as in the thick disk MHD
simulations of Gammie et al. (2004) where $\hat a \approx 0.93$, then the
accretion efficiency will be lower as well, $\approx 17$ \% for $\hat a
\approx 0.93$. Even in the
conservative case where accretion is via a geometrically thick disk
(and hence the spin/disk alignment is relatively inefficient) and the
initial orientation between the hole's spin and the disk rotation axis
is assumed to be random, we find that most MBHs rotate rapidly with
spin parameters $\hat a >0.8$ and accretion efficiencies
$\epsilon>$12\%.  As recently shown by Yu
\& Tremaine (2002), Elvis, Risaliti, \& Zamorani (2002), and Marconi
\etal (2004), a direct comparison between the local MBH mass density
and the mass density accreted by luminous quasars shows that quasars
have a mass-to-energy conversion efficiency $\epsilon\gta 0.1$ (a
simple and elegant argument originally provided by Soltan 1982).  This
high average accretion efficiency may suggest rapidly rotating Kerr
holes, in agreement with our findings.

In our models, there is a weak trend for the most massive and the
least massive holes to have slightly lower spin parameters
(efficiencies), the former because they experience more binary
coalescences during their lifetime, and the latter because they
experience a smaller number of accretion episodes.

One way to avoid rapid rotation and produce slowly rotating holes is
to assume ``chaotic feeding'' in which small amounts of material, with
$\Delta m\ll m_{\rm BH}$, are swallowed by the hole in successive
accretion episodes with random orientations (e.g., Moderski and Sikora
1996).  The constraints on such a model appear, however, to be stringent.
Specifically, for a thin disk the angular momentum of the inflowing
material must vary on a timescale $\lsim t_{\rm align} \ll t_{\rm
Salpeter}$.  Otherwise alignment of the hole with the outer disk
ensures that most of the mass accreted by the hole spins it up.  The
required timescale for angular momentum reversals is much shorter than
that expected in merger-initiated nuclear activity, but could in
principle be provided by (say) accretion of molecular clouds on random
orbits.

In this context, an issue which must be investigated further is the
alignment of the BH spin axis and the angular momentum axis of the
outer disk when the initial angle is more than 90 degrees. In this
case the inner disk ends up ``anti-aligned" instead of ``aligned"; to
fully align the hole with the outer disk requires the inner disk to
``flip" by 180 degrees.  If the timescale for alignment is as short as
we have assumed for the thin disk geometry, then the initial period of
``anti-alignment" is probably not important, but if the timescale is
of order the Salpeter timescale successive accretion events could
cancel out.

To explain the dichotomy between radio-quiet and radio-loud quasars,
Wilson \& Colbert (1995) suggested that in all radio-quiet quasars the
MBH is slowly rotating.  Since most holes rotate rapidly in our
models, our results suggest instead that the spin of a MBH is not a
necessary and sufficient condition for producing a radio-loud quasar.
Independent evidence for this conclusion comes from observations of
X-ray binaries, where jets are observed to be present only in the
hard/low X-ray spectral state, but not in the soft/high state (Fender
2001). This favors a model in which the mode of accretion, perhaps
together with the spin, determines the radio loudness of an accreting
black hole (e.g., Rees et al. 1982; Meier 2001).

\acknowledgements Support for this work was provided by NASA grants
NAG5-11513 and NNG04GK85G, and by NSF grant AST-0205738 (PM).  EQ is
supported in part by NSF grant AST 0206006, NASA grant NAG5-12043, an
Alfred P. Sloan Fellowship, and the David and Lucile Packard
Foundation.

{}

\end{document}